\documentstyle[12pt,amssymb,a4]{article}
\newcommand{\Gm}{\Lambda}
\newcommand{\pa}{\partial}
\newcommand{\la}{\langle}
\newcommand{\ra}{\rangle}
\newcommand{\be}{\begin{equation}}
\newcommand{\ee}{\end{equation}}
\newcommand{\bt}{\beta}
\newcommand{\gm}{\gamma}
\newcommand{\th}{\theta}
\newcommand{\ep}{\epsilon}

\newcommand{\et}{\eta}
\newcommand{\vp}{\varphi}
\newcommand{\kp}{\kappa}

\begin{document}
\begin{titlepage}
\begin{flushright}
{\rm
{\tt hep-ph/0107322}\\
July 2001\\
}
\end{flushright}
\vskip 2cm
\begin{center}
{\Large\bf Universal Singlets, Supergravity and Inflation}

\vskip 1cm
{\large
Lotfi Boubekeur \footnote{\tt lotfi@he.sissa.it}
and Gianmassimo Tasinato \footnote{\tt tasinato@he.sissa.it}
\\}
\vskip 0.4cm
{\small\it  SISSA-ISAS,  Via Beirut 4, I-34013 Trieste, Italy.\\
and\\
INFN Sezione di Trieste, Trieste, Italy. }
\end{center}
\vskip .5cm

\begin{abstract}
In supersymmetric theories, the occurrence of universal singlets is a delicate issue, because they usually induce tadpoles that destabilize the hierarchy. We study the effects of these tadpoles in supersymmetric hybrid inflation models. The resulting scenario is generically modified, but it is still possible to achieve inflation in a natural way. It is  argued that singlets, despite the problems associated with their presence, can lead to interesting cosmological consequences.

\end{abstract}
\hspace{0.9cm}
\end{titlepage}

\setcounter{footnote}{0}
\setcounter{page}{1}
\setcounter{section}{0}
\setcounter{subsection}{0}
\setcounter{subsubsection}{0}


\newpage

\section{Introduction}
In Particle Physics, the introduction of singlet fields has been invoked in many models to solve various problems. This is done for instance in the Standard Model to give masses to neutrinos with the see-saw mechanism, or in the so called NMSSM for other purposes. In  other  cases, their presence is actually unavoidable, like in theories that require compactification from higher dimensions.
 However, it has been pointed out that the presence of these fields  induces generically new quadratic divergences at one (or more) loop(s), in particular tadpoles (terms linear in the singlet) that destabilize dramatically the hierarchy \cite{old, BP, BPR}. Some efforts have been done to show how to 'tame' these divergences in supergravity, exploiting them to solve some notorious problems \cite{Kolda, KPP}.

Also in Cosmology, singlets have been shown to be very useful.  For example, it was pointed out that singlets can be useful to provide a strong first order phase transition essential for a successful baryogenesis in the NMSSM \cite{singlets}. In some inflationary models their presence, even if less stressed, is required. However, the tadpole contributions have never been taken into account in the cosmological context. Due to their particular properties, singlets are sensitive to the Planck scale physics. Since Cosmology is the study of the early stages of the universe (just after the Planck era), it is perfectly legitimate to ask whether their presence lead to some consequences. In this paper, we will consider the modifications required by the presence of these tadpoles in the hybrid inflationary scenario.

By now, it is well established that the inflationary paradigm provides a successful and elegant solution to three essential questions  of standard Cosmology: the horizon, the flatness and the monopole problem\cite{Guth, Linde}.
It is also widely hoped that successful inflationary models could emerge naturally from pure Particle Physics considerations \cite{Lyth-Riotto, Dvali-Riotto}, in the sense that any consistent particle model may have a built-in sector that ensures inflation. Supersymmetric hybrid inflation models appear to be the most promising to achieve this task. Such models (and their extensions) have been constructed and studied extensively \cite{Lazarides}. Typically, they are based on superpotentials of the form $W_{\rm inflation}=\kp S (\Phi \bar{\Phi}-\mu^2)$,  where $\kp$ is a dimensionless coupling constant, $S$ is a singlet superfield and $\Phi$, $\bar{\Phi}$ are superfields that are conjugate under some non trivial representation of a group $G$. At a certain time, inflation is dominated by the $F$-term of the singlet field ($V_0=\mu^4$), and this explains the presence of the linear term in the previous superpotential. Usually $\Phi$ and $\bar{\Phi}$ are taken to be the Higgs fields that break the GUT gauge symmetry so that  $\mu\sim M_{\rm GUT}$. The resulting scalar potential is the prototype of hybrid inflation \cite{hybrid} except for the mass term for $S$, which is essential to drive the inflaton to its minimum. Such a slope can however be generated, independently from soft breaking mass terms, by the one loop corrections to the scalar potential along the inflationary trajectory \cite{Dvali-Shafi}. This model succeed in reproducing the correct values of density perturbation and the spectral index at the price of a small coupling constant ($\kp\sim 10^{-3}-10^{-4}$). The generic problem of inflationary models is the stability of the potential. In other words: how to keep the inflaton potential flat enough to achieve successful inflation? 
Generally, without $D$-term contribution, supergravity gives new terms to the effective potential of the inflaton that usually destroy the flatness of the potential. However, it is argued that these corrections can be brought under control via a judicious choice of the K\"ahler potential and the superpotential \cite{Stewart2, Lazarides}. Models of inflation in which $D$-term contributions are considered have been studied \cite{D-inflation}, showing that it is possible to evade the problems associated with supergravity corrections (See however \cite{xi}). 

As we have seen, many characteristics of supersymmetric models have been largely used in building inflation models. In fact, the singlet nature of the inflaton is a crucial feature, since it protects it from acquiring a too large mass, that will ruin inflation.  However, the particular properties of singlets have  not been  explored yet in inflation, and this is the main purpose of this paper, at least in a specific example. We will see, in a particular model, that the presence of singlet fields provide a Particle Physics realization of a specific version of  hybrid inflation, the so called mutated hybrid inflation \cite{Stewart1}. 
\vskip 0.1 cm
The paper is organized as follows. In Section 2, we briefly review the properties of singlets in supergravity. In Section 3, we will focus our discussion on the case of the superpotential of supersymmetric hybrid inflation, showing that  the presence of tadpoles generically  changes the scalar potential that drives inflation. In Section 4, without analyzing in full details the consequences of these modifications, we notice that, in a certain regime,  the modified scalar  potential can provide a realization of the  mutated hybrid inflation scenario. Section 5 is devoted to the study of the stability of the potential under one-loop  and supergravity corrections. Finally, in Section 6, we give our conclusions.

\section{Universal Singlets in Supergravity}

In Particle Physics models, universal singlets are fields that do not transform under any  gauge symmetry of the Lagrangian. Therefore,  roughly speaking, in non supersymmetric models containing a scalar singlet field $s$,  nothing will forbid the appearance in the Lagrangian  of terms such as $ a \Gm^3 s + b \Gm^2 s^2 + c \Gm s^3 + {\rm h.c.}$ with $a$, $b$, $c \sim {\cal O}(1)$. Moreover, the natural value for $\Gm$ is $M_P$\footnote{Throughout the paper, $M_P$ stands for the reduced Planck scale, namely $M_P=M_{\rm Planck}/\sqrt{8 \pi} \simeq 2.4 \times 10^{18}$ GeV.}, so singlets will get masses and vev's of ${\cal O}(M_P)$. If not coupled to light fields, they will decouple from the low energy theory. Instead, if they are coupled to light fields, they will communicate to them their large vev, destabilizing dramatically the hierarchy. 

One could think that invoking supersymmetry will ameliorate the things, but the situation remains the same also in SUSY models \cite{old}. Indeed, it has been shown that, if a supergravity model contains singlets, they can destabilize the mass hierarchy, introducing new quadratic  contributions coming from tadpoles \cite{BP, BPR}. These new quadratic terms have been used to communicate supersymmetry breaking in a particular way \cite{Nilles}, to generate the GUT scale \cite{Kolda} and to solve the $\mu$-problem \cite{KPP} (See also \cite{GM} for an early attempt).

\vskip 0.2 cm
For concreteness, 
let us consider a  supersymmetric model with a visible sector containing an universal singlet superfield $S=s+\th^2 F_s$, and a hidden sector, whose fields are denoted generically with $\Sigma= \sigma+\th^2 F_{\Sigma}$, responsible for supersymmetry breaking. 
Following \cite{Nilles}, tadpoles arise due to terms like
\be
\label{eq:4}
\delta K=\left[ 1+\frac {c}{M_P} (S+S^{\dagger})\right]\Sigma \Sigma^{\dagger}
\ee
in the K\"ahler potential. The higher order term, proportional to $c$, is allowed by all the gauge symmetries, and it is generically present in the K\"ahler potential just because $S$ is a universal singlet.

The low-energy Lagrangian contains the following $D$-term contribution \cite{Nilles} \footnote{The expression (\ref{eq:5}) comes from a full supergravity computation, see \cite{BP, BPR} for more details.}
\be
\label{eq:5}
{\cal L}_D=\int  d^2\th  d^2\bar{\th}{\textrm \,\,\,} e^{K/M_P^2}K,  
\ee
where $K$ here is the K\"ahler potential written in terms of superfields.
After integrating out the hidden fields, the effective potential coming from the tadpole is given by \cite{Nilles, Kolda}
\be
\label{eq:6}
\Delta V_{\rm tadpole}= \gm  \frac{M_f^4}{M_P}(s+s^{\dagger})+(\alpha \bt F_s M_f^2+ {\rm h.c.} )
\ee
where $\alpha$ is a parameter (See \cite{Kolda, KPP})  related to the SUSY breaking in the hidden sector, and $\bt$ and $\gm$ are loop factors that are less than one. The mass  $M_f$ stands for the scale of breaking of supersymmetry in the hidden sector, i.e. $\la F_\Sigma \ra =M_f^2$. The loop factors and $\alpha$ will be an essential ingredient for our discussion\footnote{The values of $\alpha$, $\bt$ and $\gm$ are model-dependent. We consider them as free parameters in their respective allowed range.}. 
 They are related to $c$, to the number of hidden fields and to the detailed structure of the K\"ahler potential; their typical value is in the range ${\cal O}(1-10^{-4})$. In the usual  gravity mediated supersymmetry breaking models, one  arranges for $M_f^{2}\sim m_{3/2}M_P$, where the ``gravitino mass'' is chosen $m_{3/2} \lesssim O({\rm TeV)}$, to solve the hierarchy problem. 

The full scalar potential will include, in addition to  standard terms, the tadpole contribution (cf. Eq. (\ref{eq:6})). In terms of  auxiliary field $F_S$ it reads \cite{Kolda}

\be \label{potaux}
V_{F_S}=(\bt \alpha M_f^2 F_S + {\rm h.c.}) - |F_S|^2-\left( F_S \frac{\pa W}{\pa S}+ {\rm h.c.}   \right).
\ee
Since the auxiliary fields $F_s$ are non dynamical, they can be eliminated using their equation of motion \footnote{Notice the presence of the extra piece in the $F$-term of $s$, which is due to the tadpole; the  effect of the tadpole is to shift the vev of $F_S$ by the amount $\alpha \bt M_f^2$.}
\be \label{aux}
F_{s}^{\dagger}=-\frac{\pa W}{\pa S}+ \alpha \beta M_{f}^{2}.
\ee
At this point, to continue the discussion, we must consider a specific form of the superpotential. In the next section, we will consider the typical superpotential for supersymmetric hybrid inflation.

\section{The model}
 Within the model of the previous section, let us plug in the superpotential of supersymmetric hybrid inflation i.e.
\be
\label{eq:7}
W_{\rm inflation}=\kp S (\Phi \bar{\Phi}-\mu^2).
\ee
$\kp$ is a dimensionless coupling constant, $S$ is the singlet chiral superfield, while $\Phi$ and $\bar{\Phi}$ are chiral superfields, belonging to the visible sector, that are conjugate under a non trivial representation of some group $G$.
 One can always impose an appropriate  $R$-symmetry \footnote{These symmetries are global, they are likely to be broken by gravitational interactions, so at the end $S$ will not carry any quantum number.} such that the superpotential (\ref{eq:7}) is the most general renormalizable one. We do not specify the form of the superpotential for the hidden sector.

At tree level, the scalar potential is readily computed. It is 
\be
V(\vp,\bar{\vp},s)=\kp^2|\vp\bar{\vp}-\mu^2|^2+\kp^2|s|^2(|\vp|^2+|\bar{\vp}|^2) + D\rm{-terms}.
\ee

where $s$, $\vp$ and $\bar{\vp}$ are the scalar components of $S$, $\Phi$ and $\bar{\Phi}$. 
We will restrict ourselves to the $D$-flat direction $|\vp|=|\bar{\vp}|$. Minimizing the potential, one finds that there are two sets of minima. The first is the supersymmetric one, it is located at $|\vp|=\mu$ and $s=0$. The second one breaks SUSY, for $ s > s_c= \mu$ and $|\vp|=0$. 
Inflation in this scenario proceeds by assuming chaotic initial conditions for the fields 
$s$ and $\phi$. That is, the 
inflaton field $s$ rolls from $s\gg s_c$ towards the true minimum ($s=0$), while the 
"auxiliary" field $\vp$
 is held at the origin. The universe undergoes an exponential expansion phase (inflation) since its energy 
density is  then dominated by the false vacuum one ($V=\kp^2 \mu^4$). But this will not last 
forever; as soon as $s$ reaches the critical value $s_c$, 
all the fields rapidly adjust to their SUSY vacuum values restoring supersymmetry, and 
inflation finishes.

\vskip 0.1 cm
\noindent
Let us include the tadpole contributions to the scalar potential. Using Eqts. (\ref{potaux}) and (\ref{aux}), one obtains the scalar potential as a function of the two fields $\vp$ and $s$
\be
\label{V}
V=\alpha^{2}\bt^{2}M_{f}^{4}+\gamma \frac{M_{f}^{4}}{M_{P}}(s+s^{\dagger})+2 \kp^{2}|s|^{2}|\vp|^{2}-2\kp\alpha\bt M_f^2(|\vp|^2-\mu^2)+\kp^2(|\vp|^2-\mu^2)^2.
\ee
Clearly, due to the presence of the linear term in $s$, the minimum for $s$ is no more at the origin, but it is now given by

\be
s=-\frac{\gm M_f^4}{2 \kp^{2} M_P |\vp|^2}.
\ee

The supersymmetric minimum is recovered when $\gamma=0$. This corresponds to choose $c$ exactly zero in the expression of the K\"ahler  potential (\ref{eq:4}). However, a priori, we have no obvious reason to enforce it to this value.

The result is that 
the values of $s$ and $|\vp|$ are now correlated, and while $s$ rolls down along the inflationary trajectory, $\vp$ moves away from the origin. The usual scenario for hybrid inflation is modified, but  the new characteristics of the model can still be used in an inflationary context. For simplicity, we will set the scale $\mu$ to zero in the scalar potential. The scale $M_{f}$, in our case, can take any value below the Planck scale ($M_f\le M_P$), since we do not aim to provide  a phenomenologically acceptable scenario for supersymmetry breaking.  We imagine that this is achieved by some other sector of the model.

The resulting potential, with $\mu$ set to zero, looks similar to another realization of hybrid inflation, the mutated hybrid inflation. Indeed, some years ago, Stewart proposed a new version of hybrid inflation based on a potential of the form \cite{Stewart1}
\be
\label{eq:2}
V(\phi,\psi)=V_0\left(1-\frac{\psi}{M} \right)+\frac{\lambda}{2} \psi^2 \phi^2.
\ee
The inflationary trajectory is obtained by minimizing on $\psi$. Along this trajectory, both $\psi$ and $\vp$ roll. The potential, as a function of $\vp$,  reads
\be
\label{eq:3}
V=V_0 \left( 1-\frac{V_0}{2 \lambda M^2 \phi^2}\right).
\ee
Stewart argued that such a potential can arise from an effective superpotential due to non perturbative effects such as gaugino condensation. In the next two sections, we will see that the addition of singlet tadpoles will provide a new particle physics motivation to this model.

\section{Inflating with tadpoles}
Let us proceed to analyze our potential. Minimizing with respect to $s$, we end  with the scalar potential for the inflaton field $\varphi$
\be
\label{eq:8}
V=M^4_f \alpha^2 \bt^2 \left( 1-\frac{\gm^2 M_f^4}{2 \kp^2 \alpha^2\bt^2|\vp|^2M_P^2}\right)+\kp^2|\vp|^4-\kp\alpha\bt M_f^2(\vp^2+\vp^{\dagger2})
\ee
The potential (\ref{eq:8}) looks very similar to the one of mutated hybrid inflation, except for the two last terms. In order to ignore them we must impose

\be
\label{eq:9}
\xi \ll\left(\frac{\bt\alpha}{\kp}\right)^{1/2}, 
\ee
where we have defined $\vp=\xi M_f$.
Furthermore their first and second derivatives must also be negligible with respect to the derivatives of the first term that is supposed to drive inflation. These requirements translate into the following condition 
\be
\label{eq:10}
\xi \ll \left( \frac{\gm M_f}{\kp^2 M_P} \right)^{1/3}.
\ee 
To satisfy the slow roll conditions
\be
\label{eq:11}
\ep=\frac{M_P^2}{2}\left( \frac{V'}{V}\right)^2 \ll 1 \textrm{ \,\,and\,\,}|\et|=M_P^2\left|\frac{V''}{V} \right| \ll 1,
\ee
we must have
\be
\label{eq:12}
\xi\gg\left(\frac{\gm}{\bt\alpha\kp} \right)^{1/2}.
\ee
The number of $e$-folds is given by
\be
\label{eq:13}
N=\frac{1}{M_P^2}\int {\rm d}\vp \frac{V}{V'}\simeq \frac{1}{4}\xi^4 \left( \frac{\bt\alpha\kp}{\gm}\right)^2.
\ee
The COBE density perturbation normalization corresponds to
\be
\label{eq:14}
\frac{V^{3/2}}{M_P{}^3 V'}\simeq 2 \sqrt{2} N^{3/4} \left(\frac{\kp}{\gm}\right)^{1/2}(\alpha 
\bt)^{3/2} \left(\frac{M_f}{M_P} \right)=6 \times 10^{-4},
\ee
and for $N \approx 60$, we obtain the following expression for $M_{f}$
\be
\label{eq:15}
M_f\simeq 10^{-5}\left(\frac{\gm}{\kp}\right)^{1/2}\frac{M_P}{(\alpha\bt)^{3/2}}.
\ee
As in the usual mutated hybrid inflation \cite{Stewart1}, the spectral index of density perturbations is given by
\be
n\simeq 1-6\ep+2\et\simeq 1- \frac{3}{2N}.
\ee
For $N \simeq 60$, it gives $n \simeq 0.975$.

Combining Eqts. (\ref{eq:15}), (\ref{eq:10}) and (\ref{eq:12}) one ends with
\be \label{eq:20}
\kp\ll 10^{-5}.
\ee
This constraint is not surprising. In fact the smallness of the coupling constant $\kp$ is a typical prediction of  hybrid inflation models \cite{Lazarides}.

Eventually, combining Eqts.(\ref{eq:13}) and (\ref{eq:15}), we obtain
\be
M_f\simeq   10^{-5} \frac{\xi}{\bt\alpha} M_P.
\ee
\noindent
To achieve inflation, the parameters of the model must obey various constraints. However, it is possible to fulfill them in a natural way. As an example, we take  $\alpha$ and $\bt$ to their maximal value i.e. $\alpha$, $\bt \sim 1$: this choice allows to avoid fine tuning for the other parameters. 
Taking $\kp \sim 10^{-6}$, one can consider the loop factor $\gamma$ in the allowed range $\gamma \sim 10^{-1}-10^{-4}$. Consequently, the range for $\xi$ is $10 \ll \xi \ll 10^{3}$.
 We get a  scale of SUSY breaking of the order $M_{f} \simeq 10^{14}-10^{17}$ GeV, and the lower one ($M_{f} \simeq 10^{14}$ GeV) is the typical scale for a model of mutated  hybrid inflation.
\vskip 0.1 cm
\noindent

Usually, inflation finishes when the slow roll conditions are no more 
valid. This happens generally before the inflaton reaches the true minimum. 
There the inflaton begins oscillating coherently reheating the universe. 
Also in our model, the inflation ends when the slow roll conditions, represented by formula (\ref{eq:12}), break down. Actually, the inflaton field  energy lies between the two scales given by equations  (\ref{eq:12}) and  (\ref{eq:10}): this means that nor the inflaton $\vp$ nor the singlet $s$ reach the true minimum of the scalar potential at the end of inflation.

\section{Stability of the potential}

\vskip 0.2 cm
\noindent 

The tree level scalar potential usually receives corrections due to loop 
effects and to supergravity contributions. Such corrections, in 
our case \footnote{In some models, these corrections are actively used to drive inflation (see as an example  \cite{Linde-Riotto}), but we will not consider this possibility.}, are dangerous because they can destabilize the inflationary trajectory. 

\vskip 0.2 cm
The one-loop corrections, as in usual superymmetric theories, depend on the 
mass splitting between the members of the supermultiplet, induced by the 
supersymmetry breaking. More precisely, the Coleman-Weinberg one-loop
effective potential \cite{CW} shows that these corrections are proportional to
the fourth power of the mass splitting. In our case, it is easy to see
that this quantity, being proportional to the tiny coupling constant $\kp$
(See Eq. (\ref{eq:20}), is small enough to render these corrections negligible
during the inflationary era.

\vskip 0.2 cm
Unfortunately, the situation with supergravity corrections is much more 
delicate. Although tadpole  contributions, which are an essential ingredient for our 
model, come from a $D$-term, our scenario is actually an $F$-term inflationary one.
 Consequently the scalar potential receives the usual supergravity 
corrections to $F$-terms.

As  clearly explained in \cite{Stewart2}, these corrections are generically 
non negligible \footnote{Unless some fine tuning in the K\"ahler potential is made either by choosing the arbitrary K\"ahler couplings to be very small \cite{Lazarides} or by choosing a specific form of the K\"ahler potential (and the superpotential), that can be ascribed for example to superstrings constructions \cite{Stewart2}.},
 and one should expect new contributions to  the 
scalar potential in Eq. (\ref{V}), proportional to 
$M_{f}^{4}(|\vp|^{2}+|s|^2)/M_{P}^{2}$. In our case, due to the fact that 
the scale $M_{f}$ is so large, these corrections are potentially important. 
Hopefully, other contributions, in a more refined version of our model, would cancel or keep under control such dangerous terms. However we will not consider this issue since it is out of the scope of the paper (See \cite{sugra1, sugra2, sugra3} for interesting ideas in this direction).
\section{Conclusions}

The presence of singlets in supergravity is a problematic issue, because they usually destabilize the hierarchy. Only in the past few years, it has been realized that their properties can provide interesting phenomenological models in Particle Physics \cite{pheno}. Singlet fields, in the past, have also been used in Cosmology. For example, it was pointed out in \cite{singlets} that singlets can be useful to provide a strong first order phase transition essential for a successful baryogenesis in the NMSSM, and moreover they are extensively used in inflationary models. 

In this paper, we have shown that these fields can have other cosmological applications, and  in a supergravity framework. 
Indeed, we have shown that due to the presence of the tadpole contributions, the usual hybrid inflation scenario is generically modified. We point out that it is possible to use singlet tadpoles in a simple way  to provide a new realization for a different  scenario of hybrid inflation: the so called mutated hybrid inflation. In this framework, we have shown that it is possible to obtain an inflationary regime for a natural choice of the parameters. 

There is no doubt, despite the unavoidable problems associated to their presence, that singlets tadpoles can lead to interesting cosmological implications.
\section*{Acknowledgements}

We thank R. Jeannerot, D. H. Lyth, A. Masiero,  A. Riotto and L. Sorbo for useful discussions. We are also especially grateful to L. Covi for interesting discussions and useful comments on the manuscript.   LB thanks G. Senjanovi\'c for his encouragement. The work of both authors was partially supported by the European RTN project `` Physics Across The Present Energy Frontier: Probing The Origin Of Mass'' under contract number HPRN-CT-2000-00148. The work of GT was also partially supported by the European RTN project ``Supersymmetry And The Early Universe'' under contract number HPRN-CT-2000-00152.

\end{document}